\documentclass[aps,prc,twocolumn,floatfix,12pts,superscriptaddress]{revtex4}
\usepackage{epsfig}
\usepackage{amssymb}
\usepackage{amsmath}
\usepackage{color}
\usepackage{graphicx}
\usepackage{epsfig}
\usepackage{amsmath}
\usepackage{color}
\usepackage{graphicx}
\usepackage{float}
\definecolor{blue}{rgb}{0.05, 0.05, 0.5}
\def \beq{\begin{equation}}
\def \eeq{\end{equation}}
\def \beqa{\begin{eqnarray}}
\def \eeqa{\end{eqnarray}}

\usepackage{xspace}

\begin{document}
\title{Thermal photons as a sensitive probe of $\alpha$-cluster in C+Au collisions at the BNL Relativistic Heavy Ion Collider}
\author{Pingal Dasgupta}
\affiliation{Key Laboratory of Nuclear Physics and Ion-beam Application (MOE), Institute of Modern Physics, Fudan University, Shanghai 200433, China}
\author{Guo-Liang Ma}
\email[]{glma@fudan.edu.cn}
\affiliation{Key Laboratory of Nuclear Physics and Ion-beam Application (MOE), Institute of Modern Physics, Fudan University, Shanghai 200433, China}
\author{Rupa Chatterjee}
\email[]{rupa@vecc.gov.in}
\affiliation{Variable Energy Cyclotron Centre, HBNI, 1/AF, Bidhan Nagar, Kolkata-700064, India}
\author{Li Yan}
\email[]{cliyan@fudan.edu.cn}
\affiliation{Key Laboratory of Nuclear Physics and Ion-beam Application (MOE), Institute of Modern Physics, Fudan University, Shanghai 200433, China}
\author{Song Zhang}
\affiliation{Key Laboratory of Nuclear Physics and Ion-beam Application (MOE), Institute of Modern Physics, Fudan University, Shanghai 200433, China}
\author{Yu-Gang Ma}
\email[]{mayugang@fudan.edu.cn}
\affiliation{Key Laboratory of Nuclear Physics and Ion-beam Application (MOE), Institute of Modern Physics, Fudan University, Shanghai 200433, China}

\begin{abstract}
Different orientations of $\alpha$-clustered carbon nuclei colliding with heavy ions can result in a large variation in the value of anisotropic flow. Thus, photon flow observables from clustered ${\rm^{12}C}$ and ${\rm^{197}Au}$ collisions could be a potential probe to study the `direct photon puzzle'. We calculate the transverse momentum spectra and anisotropic flow  coefficients ($v_n$) of thermal photons from collisions of triangular $\alpha$-clustered  carbon  and gold  at $\sqrt{s_{\rm NN}}=200$ GeV at RHIC using a hydrodynamic model framework and compare the results with those obtained from unclustered carbon and gold collisions. The  slope of the thermal photon spectra is found to vary moderately for  different orientations of collisions. However, we find that the elliptic ($v_2$) and triangular flow ($v_3$) coefficients of direct photons for specific configurations are significantly larger and predominantly formed by the QGP radiation. A strong  anti-correlation between initial spatial ellipticity and triangularity is observed in an event-by-event framework of $\alpha$-clustered ${\rm C+Au}$ collisions. These special features provide us an opportunity to detect the exotic nature of cluster structure inside carbon nucleus using the photon probe in the future experiments.
\end{abstract}

\maketitle

\section{Introduction} 
The experimental study of relativistic heavy-ion collisions provides a strong evidence of the formation of hot and dense Quark-Gluon Plasma (QGP) state of matter at the center of the collision zone~\cite{Gardim:2019xjs,Adler:2003kt,Alver:2010gr,Aamodt:2010pa,Chen:2018tnh}.  The relativistic hydrodynamic models present a satisfactory description of fireball evolution as the copious production and large harmonic flow of hadrons have been successfully explained by these frameworks~\cite{Kolb:2003dz, Huovinen:2003fa,Kolb:2000sd,Teaney:2001av,Huovinen:2006jp,Romatschke:2007mq,Teaney:2009qa,Holopainen:2010gz,Schenke:2012wb,Heinz:2013bua,ColemanSmith:2012ka, Sorensen:2010zq,Takahashi:2009na,Yan:2017ivm}.  The past studies have shown that nuclear deformation  can be investigated from the anisotropic flow of produced hadrons at intermediate and high collision energies ~\cite{Heinz:2004ir,Rybczynski:2012av,Shou:2014eya,Cao:2010bc}.  The nuclear collisions at such energies provide us a unique way to investigate the intrinsic substructure of nuclei as the time during which the participating nuclei pass through each other is too small for any slower nuclear excitation to occur and thus one can obtain a projection of overlying nuclear distributions on the fireball.

For example, the small systems, such as ${\rm p + p}$ and ${\rm p + A}$ have gathered much attention for producing  anisotropic flow coefficients comparable to the peripheral heavy ion collisions  in recent times~\cite{Heinz:2019dbd}. The measurements of ${\rm p+Au}$, ${\rm d+Au}$, and ${\rm ^3He+Au}$ collisions at 200A GeV have clearly shown the footprints of initial geometries (initial ellipticity and triangularity) on the anisotropic flow coefficients~\cite{PHENIX:2018lia}. The successful hydrodynamical description of those data has strongly suggested that the initial nucleon-level geometries have been transferred into the momentum anisotropies through hydrodynamical evolution of small droplets of the QGP in these small systems.

An analogous initial geometry effect of nuclei is also found to be present in the intermediate energy heavy-ion collisions however at a cluster level~\cite{Liu:2018eyg,Ebran}. A few recent articles have argued that the geometric $\alpha$-clustering structure of light nuclei can be realized in the realm of relativistic collisions~\cite{Broniowski:2013dia,Rybczynski:2017nrx,Bozek:2014cva}. Gamow first proposed the possibility of clustered states in light nuclei~\cite{gamow}. The identification of the fusion reaction mechanism for carbon production has led to the discovery of the clustered states~\cite{Hoyle:1954zz,Cook:1957zz}. However, many theoretical models such as no-core shell model~\cite{Roth:2011vt}, fermionic molecular dynamics~\cite{Chernykh:2007zz}, the variational Green’s function~\cite{Pieper:2002ne},  Bose-Einstein Condensate (BEC) \cite{Tohsaki:2001an,Funaki:2003af,Funaki:2006gt}, etc. have not been successful in explaining such clustered structures of nuclei.  The $\alpha$-clustering inside a light nucleus produces nuclear deformity and  spatial correlations between clusters. It has been suggested that $\alpha$-clustering configurations can be identified by giant dipole resonance~\cite{He:2014iqa,He:2016cwt} or photonuclear reactions~\cite{Huang:2017ysr,Huang:2017ypi,Huang:2020iil}. The strategy of investigating initial nucleon-level geometry via measuring flow in a relativistic heavy-ion collision has a great potential to identify the  geometric structures of clustered nuclei ~\cite{Broniowski:2013dia,Bozek:2014cva,Zhang:2017xda,Zhang:2018zzu}.

On the other hand, the direct photons are considered as one of the efficient probes to study the properties of hot and dense QGP state of matter~\cite{Srivastava:2008es,Gale:2014dfa, Chatterjee:2013naa, Shen:2013cca, Monnai:2014kqa, McLerran:2014hza, Liu:2012ax, Basar:2012bp, Tuchin:2012mf, Zakharov:2016mmc, Vujanovic:2014xva,Li:2010ts,Ma:2013bia,Adare:2008ab,Adam:2015lda}. The  thermal photon spectra obtained from the hydrodynamical model evolution combined with the prompt photon contribution from initial partonic hard scatterings have satisfactorily described the measured direct photon spectra in the range $p_T > 2$ GeV in heavy-ion collisions at both Relativistic Heavy Ion Collider (RHIC) and the Large Hadron Collider (LHC). The theoretical calculation of spectra in the region $p_T<2$ GeV has, however, been found to underestimate the data where the contribution from the hadronic phase is significant. The anisotropic flow of direct photons is also believed to be a potential probe for illustrating the initial state of a collision. However, the hydrodynamic model calculations of photon production failed to describe the elliptic  and triangular flow data at RHIC and LHC by a large margin, which has been addressed as the ``direct photon puzzle"~\cite{Adare:2015lcd,Acharya:2018bdy}. Several studies using a more realistic hydrodynamic framework, initial conditions, the pre-equilibrium flow, and holographic thermal-photon emission rate have shown improvements in the theoretical prediction but could not resolve the puzzle yet~\cite{Dasgupta:2017fns,Shen:2015nto,Gale:2020xlg,Iatrakis:2016ugz}. Recent studies have shown that the spectra and  anisotropic flow coefficients of photons  (i.e., $v_{1},v_{2},v_{3}$) from different collision systems and their simultaneous comparison could be valuable to understand the sources of the puzzle~\cite{Chatterjee:2017akg,Dasgupta:2016qkq,Dasgupta:2019whr}. A significant improvement in the photon anisotropic flow has been reported recently by  Naboka {\it et al.} using integrated hydrokinetic model~\cite{Naboka:2017dnn,Naboka:2019vko}.

In the present article, we aim to study the intrinsic geometry of the $\alpha$-clustered ${\rm C}$ nucleus using the photon probe. We calculate the spectra and anisotropic flow coefficients $v_n$($n=2,3$) of thermal photons in (2+1) dimensional inviscid hydrodynamic framework from the $\alpha$-clustered ${\rm C+Au}$ collisions at the center-of-mass energy $\sqrt{s_{\rm NN}}$=200 GeV.  The same set of observables from the unclustered ${\rm C+Au}$ collisions (see Section II) is also calculated for comparison.

This paper is organized as follows. In the next section, we briefly discuss the initial parameters and the framework for the model calculation. In section III we present how initial spatial anisotropies  $\varepsilon_{n}$ and the momentum anisotropies of thermal photons $v_n$ are calculated. In section IV we discuss the results of thermal photon spectra and anisotropic flow coefficients and finally, we summarize our results in section V.

\section{Initial condition and hydrodynamic framework}
  In the present work, we adapt the  same procedures followed in~\cite{Bozek:2014cva} to choose the initial parameters for the nuclear density distribution of $\alpha$-clustered carbon. We study different collision scenarios incorporating two types of carbon nucleus in the Monte Carlo Glauber (MCG) framework : 
  \begin{figure}[tb]
  	\centerline{\includegraphics*[width=8.8cm,clip=true]{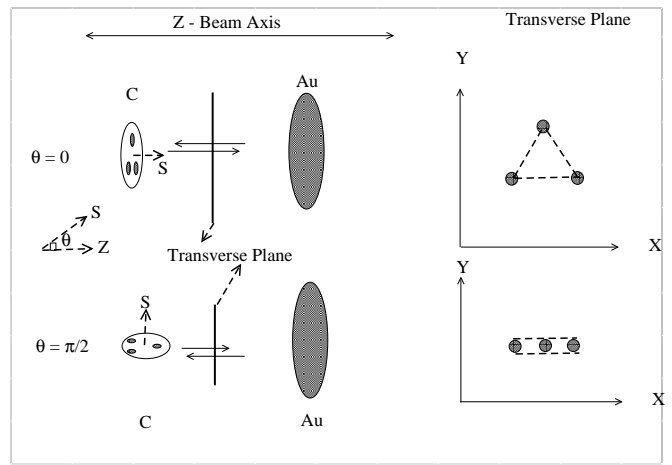}}
  	\caption{{A schematic view of collisions} of the $\alpha$-clustered carbon with a gold nucleus. The different orientations of clustered-carbon at the time of impact lead to different initial geometries (left). The triangular and line geometry appear when clustered carbon hits the gold wall with an angle 0 and $\pi$/2 respectively (right). The angle is measured between the beam axis (z-axis) and the perpendicular axis of the triangular carbon nucleus in $z-y$ plane.}
  	\label{fig1}
  \end{figure}
  
\begin {itemize}
\item {\bf $\alpha$-clustered carbon}: This is an equilateral triangle shaped $\alpha$-clustered state of carbon. It is considered as the ground state of clustered phase with binding energy $7.17$ MeV/nucleon. There exists another possible excited state of the clustered structure, a chain-like structure of 3 alpha particles with binding energy $7.47$ MeV/nucleon. However, in this article, we emphasize only on the triangle shaped carbon. We use a parametric form of nucleon distribution function for each $\alpha$ cluster \cite{Bozek:2014cva} and centers of such three $\alpha$ clusters are placed on the vertices of an equilateral triangle. The parameters are fixed in such a manner that one-body radial density distribution of the center of nucleons obeys the results of the BEC model~\cite{Funaki:2006gt}. 
The distribution of nucleons in each cluster is expressed through a Gaussian function as follows :
\begin{eqnarray}
f_i(\vec{r})=A \exp \left (- \frac{3}{2} \, (\vec{r}-\vec{c_i})^2/r_\alpha^2 
\right ), 
\label{alpha_dist}
\end{eqnarray}
where, $\vec{c_i} $ represents the center of the $i^{th}$ cluster. $r_\alpha$ determines the mean radius of each cluster.  We generate positions of nucleons in each alpha cluster from the above distribution function and then shift three of such clusters to the vertices of a triangle of length $l$. The short distance repulsion of nucleons is also an essential criterion for light nuclei, where the distance between two centers of an N-N pair cannot be less than 0.9 fm~\cite{Bozek:2014cva}.   The nucleonic positions are finally shifted to a new coordinated system where the center-of-mass resides on the origin. The parameters $l$ and $r_\alpha$ in Eq.(\ref{alpha_dist}) are  taken as $3.05$ fm and $0.96$ fm  respectively to match the results from BEC model calculations~\cite{Bozek:2014cva}. 
\item {\bf Unclustered carbon}: This is a mean-field state of ${\rm C}$ nucleus. The nuclear distribution is isotropic for this  case.   A two-parameter Wood-Saxon distribution of nuclear density is considered for an unclustered carbon where the parameters are adjusted in such a way that the root mean square radius of the nuclear distribution remains the same as the clustered carbon ($\sim 2.26$ fm).
\end{itemize}
To distribute the nucleons inside a gold nucleus we employ the standard two-parameter Wood-Saxon nuclear density distribution profile.  We assume that the beam axis is stretched along  the $z$-direction and the impact parameter lies along  the $x$-direction in each event of the collision. 

The conventional two-component MCG technique is then used to distribute initial entropy density on the transverse  plane.  The criterion for a collision between the two incoming nucleons is made as $d^2 < \frac{\rm \sigma_{NN}}{\pi}$, where $d$ is the transverse distance between the colliding nucleons and { $\rm \sigma_{NN}$} is the inelastic nucleon-nucleon cross-section which at RHIC is about 42 mb.   A participant is given a weight of $(1-\nu)$ and a binary collision is given a weight of $\nu$, where 
$\nu$ is taken as 0.145~\cite{Bozek:2014cva}. The entropy density distribution, $s(x,y)$, is obtained by  taking a weighted sum over all the sources as explained in Eq.(\ref{imple_eqn}) below.
\begin{eqnarray}
s(x,y)=K\sum_{i,j=1}^{N_{\rm part},N_{\rm coll}} [  \, \nu \, n_{\text{coll}}(x_i,y_i) f_{i}(x,y)
\nonumber \\ +(1-\nu)\, n_{\text{part}}(x_j,y_j)\ f_{j}(x,y) ],
\label{imple_eqn}
\end{eqnarray}
where, $n_{\rm part}$ and $n_{\rm coll}$ denote the number of participant and binary collision sources at the $(x,y)$ position respectively. K is a normalization factor which decides the total multiplicity of an event. The function $f_{i,j}(x,y)$ is a normalized distribution of the following form centering  about the {$i^{th}$ collision  or $j^{th}$ participant} source :
\begin{equation}
\label{normal_dist_eq}
f_{i,j}(x,y)=\frac{1}{2\pi \sigma ^2}\, e^{ -\frac{(x-x_{i,j})^2+(y-y_{i,j})^2}{2\sigma ^2}}.  
\end{equation}
The initial entropy density profile which serves as an input for our hydrodynamic calculation is obtained by taking an initial state average of $ 10000$ events with random nucleonic positions at impact parameter $b \approx 0$ fm. We modify the (2+1) dimensional ideal relativistic hydrodynamic model~\cite{Holopainen:2010gz}  to obtain the space-time evolution of the above initial entropy distribution at the mid-rapidity. The initial flow velocity component, $v_x$ and $v_y$ are taken as zero. For the sake of simplicity, we take the value of initial thermalization time as $\tau_0=$ 0.17 fm/$c$ which has been used for the $\rm {Au+Au}$ collisions at RHIC~\cite{Eskola:2005tx,Chatterjee:2017akg}, although one may expect a larger value of initial thermalization time for small systems. A lattice-based equation of state~\cite{Laine:2006cp} is used and we consider a constant temperature freeze-out at 160 MeV. The same freeze-out temperature ($T_f$) is used in earlier studies of Au+Au collisions at RHIC using the same hydrodynamical model~\cite{Chatterjee:2013naa,Dasgupta:2017fns,Chatterjee:2017akg,Dasgupta:2019whr}.

\section{ANISOTROPIC FLOW OF THERMAL PHOTONS}

The production of thermal photons from the quark gluon plasma phase is estimated by using the complete next-to-leading order photon rates from Refs.~\cite{Arnold:2001ms,Ghiglieri:2013gia}. We use the parameterized rates from Ref.~\cite{Turbide:2003si} to calculate the photon production from the hot hadronic matter phase. The emission rates ($R=EdN/d^3pd^4x$) from QGP and hadronic matter phases are integrated over the space-time history to estimate the total thermal production and anisotropic flow.
The flow coefficients $v_n$ are calculated as:
\begin{eqnarray}
v_n (p_T) \ = \ \frac{\int_0^{2\pi} \, 
	d\phi\,{\rm cos} [\,n (\phi-\psi_{n})]\,\frac{dN}{p_T dp_T dy d\phi}}{\int_0^{2\pi}\,d\phi\,\frac{dN}{p_Tdp_Tdy d\phi}},
\label{v2flow}
\end{eqnarray}
where  $\phi$ is the azimuthal angle of particle's momentum and the event plane angle $\psi_n$ is determined by,
\begin{equation}
\psi_{n} = \frac{1}{n} \arctan 
\frac{\int \mathrm{d}x \mathrm{d}y \; r^2 \sin \left( n\Phi \right) \epsilon\left( x,y,\tau _{0}\right) } 
{ \int \mathrm{d}x \mathrm{d}y \; r^2 \cos \left( n\Phi \right) \epsilon\left( x,y,\tau _{0}\right)}  + \pi/n \, ,
\label{sai_nm}
\end{equation}
where $\epsilon (x,y,\tau _{0})$ is the energy density at an initial proper time $\tau_0$ at ($x,y$) point on the transverse plane, $\Phi$ and $r$ are the spatial azimuthal angle and radial distance. The  corresponding initial state anisotropy $\varepsilon_{n}$  is quantified as~\cite{Chatterjee:2017akg,Dasgupta:2019whr,Ma:2010dv}:
\begin{eqnarray}
\varepsilon_{n}=-\frac{\int \mathrm{d}x \mathrm{d}y \; r^2 \cos [\,n (\Phi-\psi_n)]    \epsilon\left( x,y,\tau _{0}\right) } 
{ \int \mathrm{d}x \mathrm{d}y \; r^2  \epsilon\left( x,y,\tau _{0}\right)}. 
\label{e_mn1}
\end{eqnarray}

\section{Results and discussions}

\subsection{Initial configurations}
\begin{figure*}[tb]
	\centering
	{\includegraphics*[scale=0.52,clip=true]{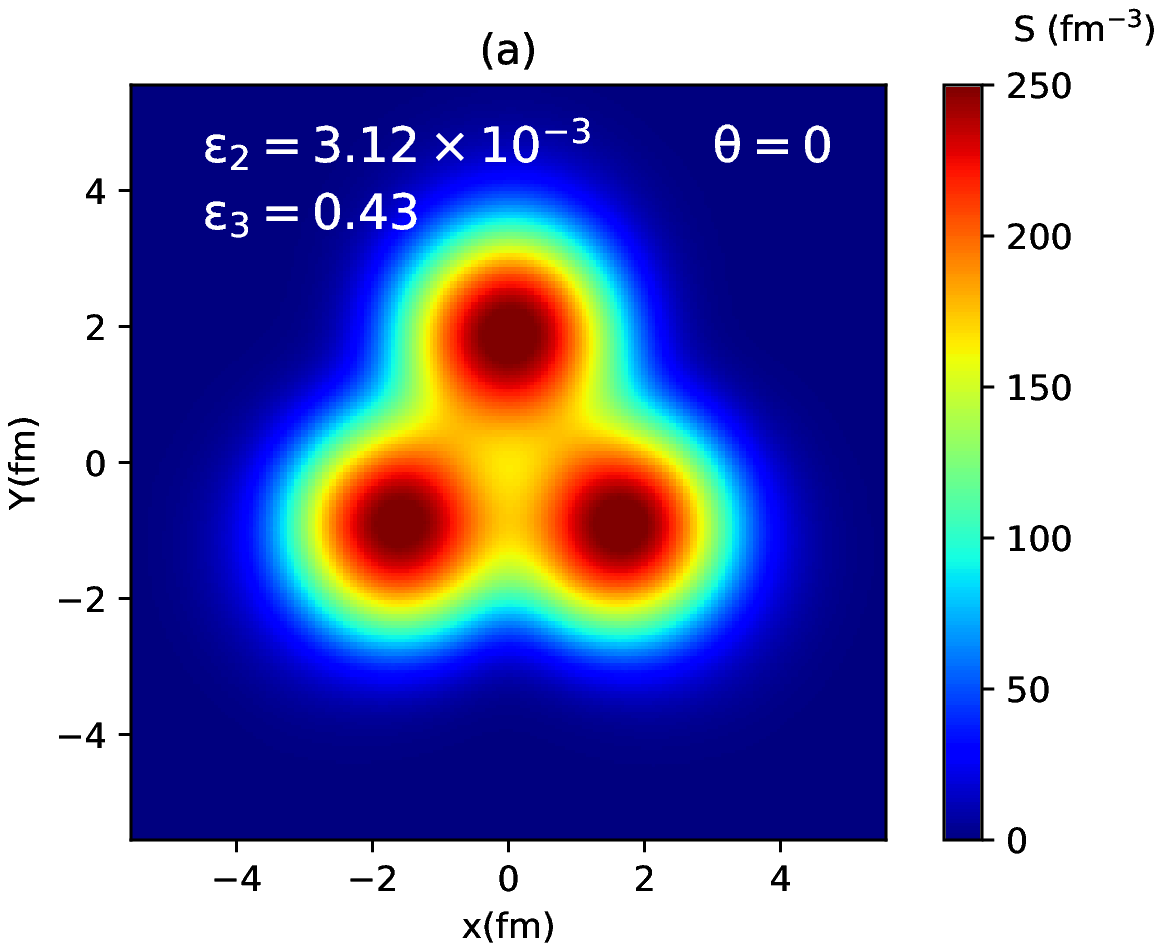}}
	{\includegraphics*[scale=0.52,clip=true]{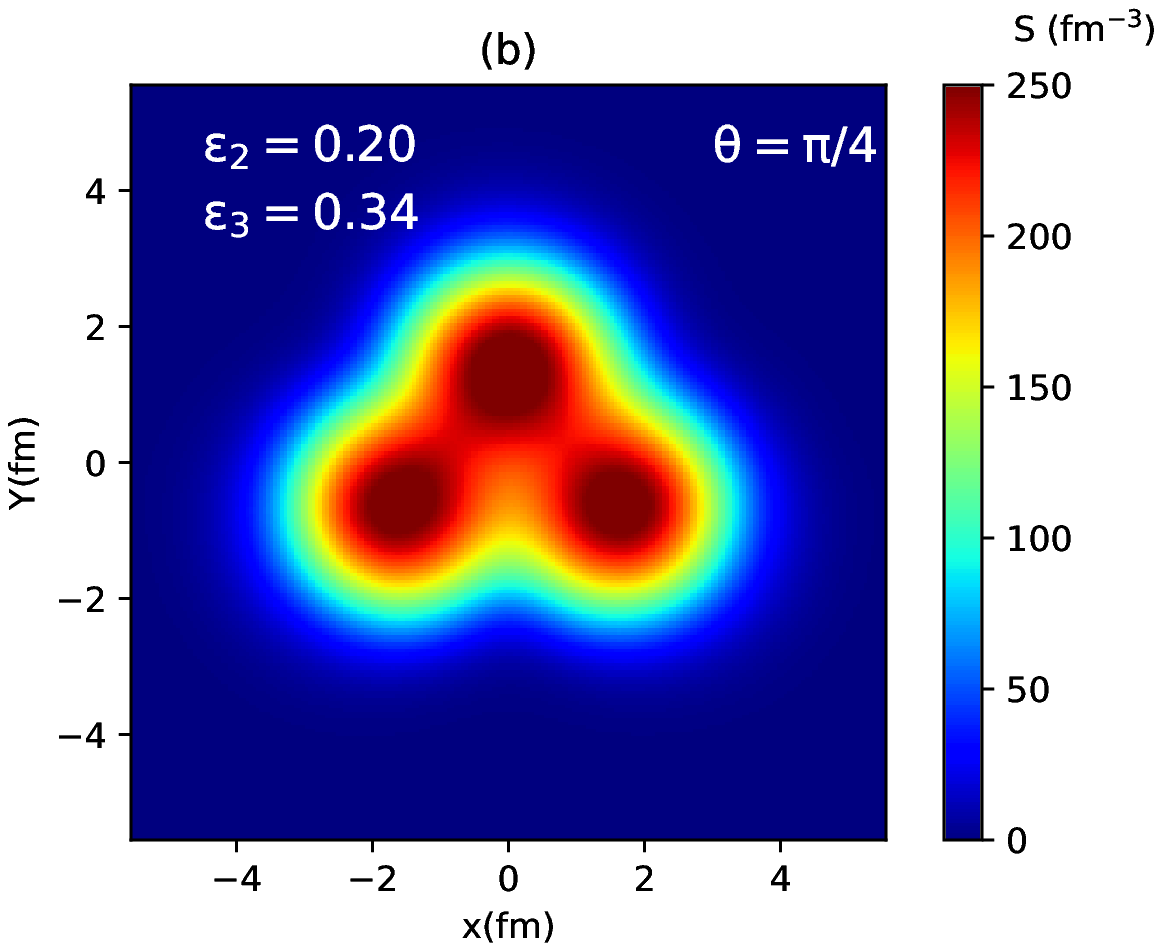}}
	{\includegraphics*[scale=0.52,clip=true]{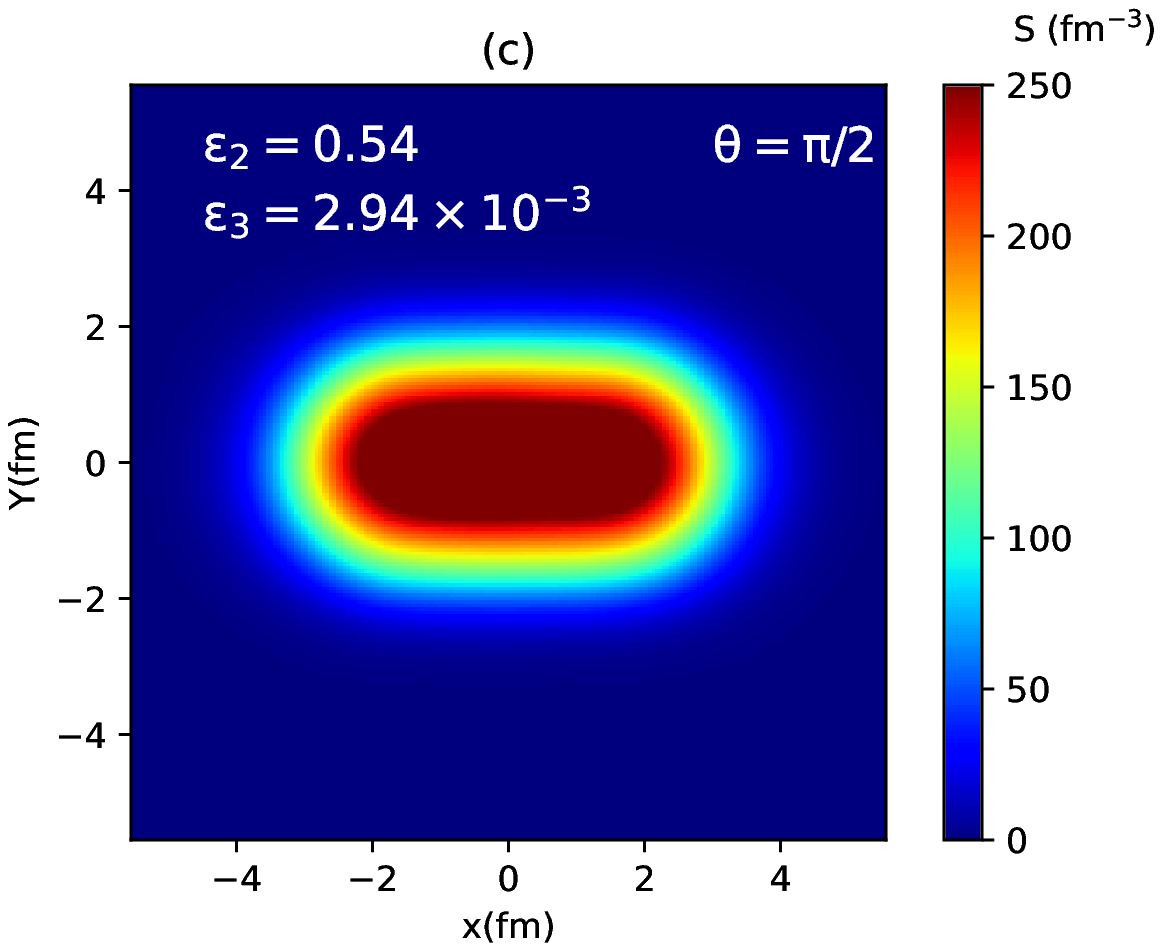}}
	{\includegraphics*[scale=0.52,clip=true]{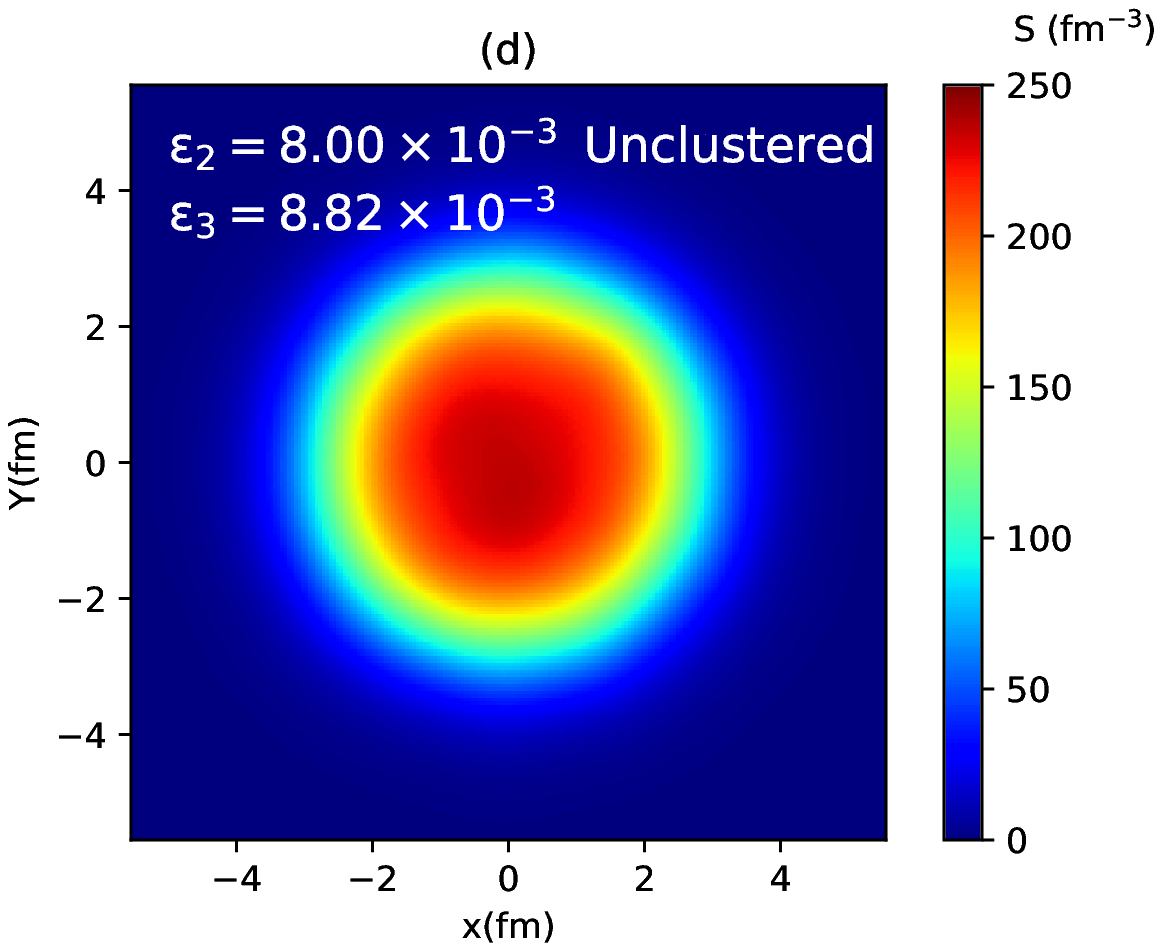}}
	\vspace{-3mm}
	\caption {(Color online) Distribution of entropy density at the formation time $\tau_0$ on transverse ($x-y$) plane for central (b $\approx$ 0 fm ) $\rm {C+Au}$ collisions at 200A GeV. The figures (a), (b), and (c) correspond to the collisions of clustered C and Au at angles $0$, $\pi/4$ and $\pi/2$ respectively, whereas,  (d) represents the initial entropy profile from  collision of unclustered C and Au. }
	\label{fig2}
\end{figure*}
\begin{figure}
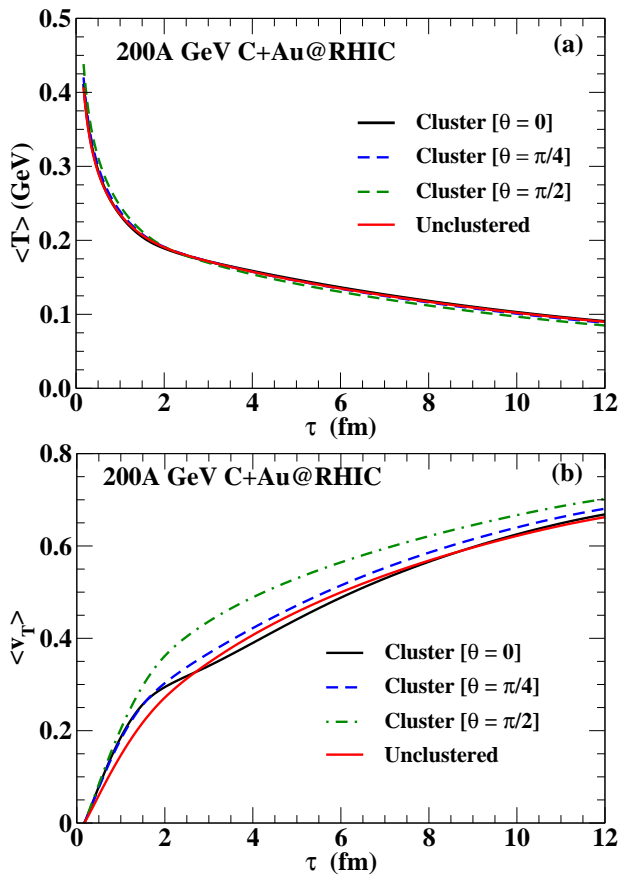

	\centerline{\includegraphics*[width=8.1cm,clip=true]{temp.eps}}
	\centerline{\includegraphics*[width=8.1cm,,clip=true]{vt.eps}}
	\caption{(Color online) (a) Time evolution of average temperature and (b) transverse flow velocity for different configurations of clustered ${\rm C}$ and ${\rm Au}$ collisions at 200A GeV. The same variables for the collisions of unclustered ${\rm C}$ and ${\rm Au}$ are shown for a comparison.}
	\label{fig3}
\end{figure}
\begin{figure}[htbp!]
	\centerline{\includegraphics*[width=8.1cm,clip=true]{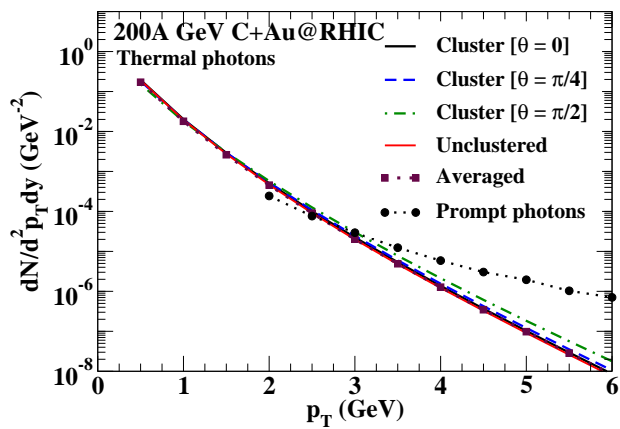}}
	\caption{Thermal photon spectra for different configurations of clustered ${\rm C}$ and ${\rm Au}$  collisions at 200A GeV. The thermal photons from the collisions of unclustered ${\rm C}$ and ${\rm Au}$  are shown for a comparison.}
	\label{fig4}
\end{figure}

 We consider three typical configurations of collisions, i.e., $\theta=0$, $\pi/4$, and $\pi/2$ in this study. This can be imagined as a deformed nucleus hitting a wall of a gold nucleus at different orientation angles (see Fig.~\ref{fig1}). Thus, the angle $\theta=0$   describes situation where the symmetry axis is aligned with the beam axis producing an initial triangular geometry in the deposited entropy distribution and the highest multiplicity amongst all orientations of collision.  For  $\theta=\pi/2$, the multiplicity is the lowest. While a  triangular geometry vanishes for such a scenario,  a large elliptic geometry appears.  We consider an intermediate situation  where the symmetry axis makes an angle $\pi/4$ with the beam axis producing both the intrinsic elliptic and triangular geometry in the deposited entropy distribution.

In Fig.~\ref{fig2} (a), (b), and (c) the event-averaged initial entropy density profile  for three orientation angles  0, $\pi$/4, and $\pi$/2 are shown respectively,  while Fig.~\ref{fig2}(d) represents the entropy profile for the unclustered carbon. We see that the entropy density at the center of each hot spot region for the clustered case is much larger than the entropy density obtained at the center of the unclustered case. In Fig.~\ref{fig2}(a) and (b), the entropy profiles are seen to have a triangular shape, while in Fig.~\ref{fig2}(c) a peanut-like structure appears with no such triangular geometry. The triangular eccentricity ($\varepsilon_3$) of the initial entropy density distribution  for the cases in Fig.~\ref{fig2}(a) and (b) is $0.43$ and $0.34$ respectively. On the other hand, the elliptic eccentricity ($\varepsilon_2$) for the cases in Figs.~\ref{fig2}(b) and (c) is $0.20$ and $0.54$ respectively, whereas $\varepsilon_2$ is vanishingly small for the other two cases.

\subsection{Evolution of temperature and flow velocity}

The time evolution of average temperature $\langle T \rangle$ and average transverse flow velocity $\langle v_T \rangle$ for the three configurations of collision for the clustered carbon and the unclustered carbon is shown in Fig.~\ref{fig3}(a) and (b) respectively. The $\langle T \rangle$ and $\langle v_T \rangle$ are obtained by Eq.~(\ref{e_mn1}) and Eq.~(\ref{e_mn2}), respectively, i.e.

\begin{eqnarray}
\langle T \rangle =\frac{\int \mathrm{d}x \mathrm{d}y \; T( x,y,\tau )   \epsilon\left( x,y,\tau \right) } 
{ \int \mathrm{d}x \mathrm{d}y   \epsilon\left( x,y,\tau \right)} ,
\label{e_mn1}
\end{eqnarray}

\begin{eqnarray}
\langle v_T \rangle =\frac{\int \mathrm{d}x \mathrm{d}y \; v_T( x,y,\tau )   \epsilon\left( x,y,\tau \right) } 
{ \int \mathrm{d}x \mathrm{d}y   \epsilon\left( x,y,\tau \right)} .
\label{e_mn2}
\end{eqnarray}
The initial average temperature (see Fig.~\ref{fig3} (a)) is found to be largest ($\sim 440$ MeV) for the configuration $\theta=\pi/2$ and smallest for the unclustered case ($\sim 407$ MeV). A relatively smaller area of the overlapping region leads to a larger $\langle T \rangle $ for the configuration $\theta=\pi/2$ at the initial time. We see an almost similar time evolution of the average temperature for all the configurations. However, the evolution of the average transverse flow velocity for different configurations are  distinct from each other (see Fig.~\ref{fig3}(b)). We see that $\langle v_T\rangle$ sharply rises for the clustered cases, whereas the rise is relatively slower for the unclustered case. The rise is maximum for the orientation angle $\pi/2$. During the time interval of $2<\tau<4$ fm/c,  we observe a sudden jerk for the configuration $\theta=0$. However, no such effect is found for $\theta = \pi/4$ and $\pi/2$. The radially outward ripples, originated at the boundary of each hotspot (see Fig.~\ref{fig2} (a) ), cause a  flow cancellation when they collide, which  effectively reduces the growth of the average transverse flow velocity. 

\subsection{Spectra and anisotropic flow}
In Fig.~\ref{fig4}, the thermal photon spectra are presented for the clustered and unclustered ${\rm C+Au}$ collisions. The thermal photon spectra for all the cases are found to be almost similar in the region $p_T< 2$ GeV. However, the spectra become sensitive to the orientation angle for $p_T>2$ GeV. The thermal spectrum for the configuration $\theta=\pi/2$ is found to be slightly above the other spectra in the larger $p_T$ region. On the other hand,  photon production from the unclustered carbon is found to be smallest. We plot a thermal photon spectrum (i.e. maroon box+dotted line) considering an orientation-averaged initial profile (averaged over 50,000 events with random orientations) for the clustered carbon and find it similar to the spectrum obtained for the unclustered carbon. We estimate the prompt photon production in the region $p_T>2$ GeV from ${\rm C+Au}$ collisions by scaling the prompt photon spectrum obtained in the p+p collisions at 200A GeV using the Monte Carlo code JETPHOX (version 1.2.2)~\cite{Aurenche:2006vj}. We expect similar production of prompt photons for collisions of clustered as well as unclustered carbon as the $N_{\rm coll}$ for both are close to each other. The Fig.~\ref{fig4} shows the prompt photons contribute significantly in the region $p_T>3$ GeV (black filled circles).

\begin{figure}[htbp!]
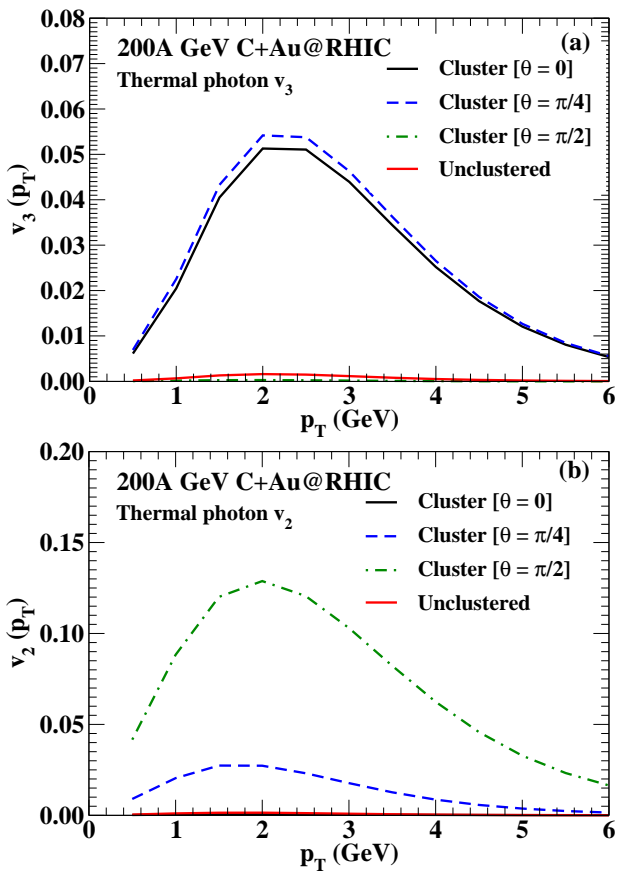

	\centerline{\includegraphics*[width=8.1cm,,clip=true]{v3.eps}}
	\centerline{\includegraphics*[width=8.1cm,,clip=true]{v2.eps}}
	\caption{(Color online) (a) Triangular and (b) elliptic flow  of thermal photons as a function of $p_T$ from different configuration of clustered ${\rm C}$ and ${\rm Au}$  collisions at 200A GeV.}
	\label{fig5}
\end{figure}

The elliptic and triangular flow of thermal photons as a function of $p_T$ are presented in Fig.~\ref{fig5}(a) and (b) respectively. We see a substantially large thermal photon $v_3$ for the configurations $\theta = \pi/4$ and $\theta = 0$. The initial triangular geometry of the entropy density profile produced in the collision of  $\alpha$-clustered carbon and gold nucleus gives rise to the triangular flow of thermal photons. As there exists no such triangular geometry for the configuration $\theta=\pi/2$ or the unclustered case, we obtain a vanishingly small  thermal photon $v_3$. The thermal photon $v_3$ for the configuration $\theta = \pi/4$ is seen to be slightly larger compared to the same obtained from the orientation angle $\theta = 0$ even though the triangular anisotropy is smaller for the former than the latter. The reason is attributed to the larger average transverse flow velocity during the evolution for the configuration $\theta = \pi/4$ than the same obtained for the configuration $\theta = 0$. On the other hand, the elliptic flow parameter of thermal photons is found to be vanishingly small for the configuration $\theta = 0$ and the unclustered case. However, the thermal photon $v_2$ is found to be very large for the orientation angle $\pi/2$. The initial elliptic eccentricity and the transverse flow velocity, both play significant role in producing a large $v_2$ of thermal photons for the configuration $\theta = \pi/2$. It is important to note that the $v_2$ obtained for the above configuration is comparable to the direct photon $v_2$ data obtained for the mid-central Au+Au collisions at 200A GeV at RHIC~\cite{Adare:2015lcd}. 

\begin{figure}[htbp!]
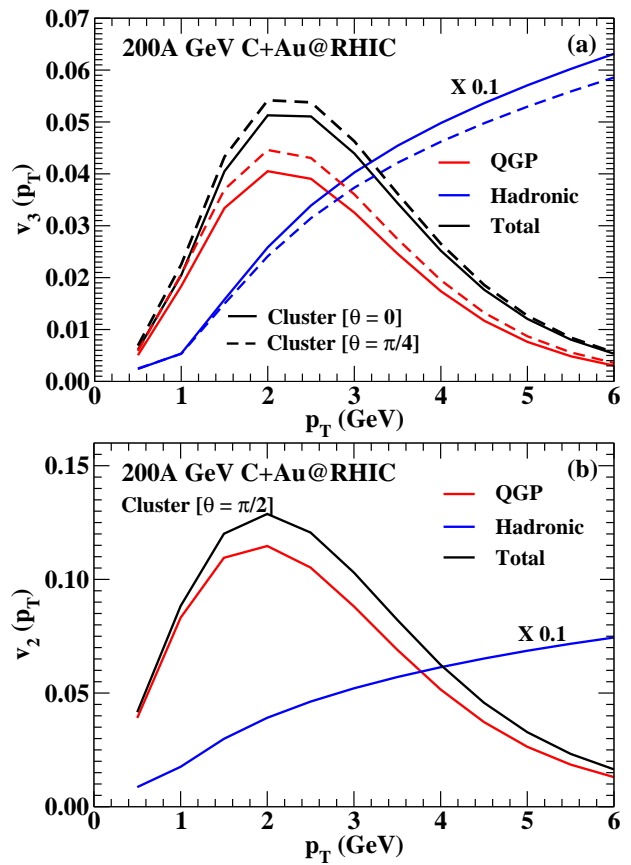

	\centerline{\includegraphics*[width=8.1cm,clip=true]{v3_qgp_had.eps}}
	\centerline{\includegraphics*[width=8.1cm,,clip=true]{v2_qgp_had.eps}}
	\caption{(Color online) The QGP and hadronic phase contributions of (a) thermal photon $v_3(p_T)$  for the orientation angle $\theta=0$ and $\pi/4$ (b)  thermal photon $v_2(p_T)$ for the orientation angle $\theta=\pi/2$. }
	\label{fig6}
\end{figure}
	
To obtain a more precise idea about the origin of the anisotropic flow coefficients of photons, we plot the individual contributions from the QGP and hadronic phases to the total value of the flow coefficients in Fig~\ref{fig6}. We calculate the contributions of thermal photon $v_3$ for the clustered case with orientation angles $\theta=0$ and $\pi/4$ in Fig~\ref{fig6}(a). We find that the maximum value of $v_3$ (at $p_T\sim2$ GeV) from the QGP phase is substantial and close to the maximum value of the total thermal photon $v_3$. Hence, the response to the initial-state triangular geometry is predominantly carried by the QGP evolution. If we compare the hadronic medium contribution to the total thermal photon $v_3$ for these cases, we find that unlike the QGP photon $v_3$, the hadronic photon $v_3$ for the orientation $\theta = 0$ is larger compared to the same obtained at the orientation $\theta = \pi/4$. The reason can be understood from Fig.~\ref{fig2}, where we see that the triangular geometry associated with  the hottest region is more prominent in Fig~\ref{fig2}(b) compared to Fig.~\ref{fig2}(a). Whereas, the boundary region, which is the source of hadronic photons, has a larger triangular geometry as shown in Fig.~\ref{fig2}(a) compared to Fig.~\ref{fig2}(b). The contributions of thermal photon $v_2$ for the clustered case with the orientation angle $\theta= \pi/2$ are plotted in Fig.~\ref{fig6}(b), and we see a similar response. The peak value of the thermal photon $v_2$ from the QGP phase is found to be about $90\%$ of the peak value of the total thermal photon $v_2$ and identically, it reflects that the most of the net thermal photon $v_2$ has built up during the early phase of QGP evolution.\\

\begin{figure}[htbp!]
	{\includegraphics*[width=7.8cm,clip=true]{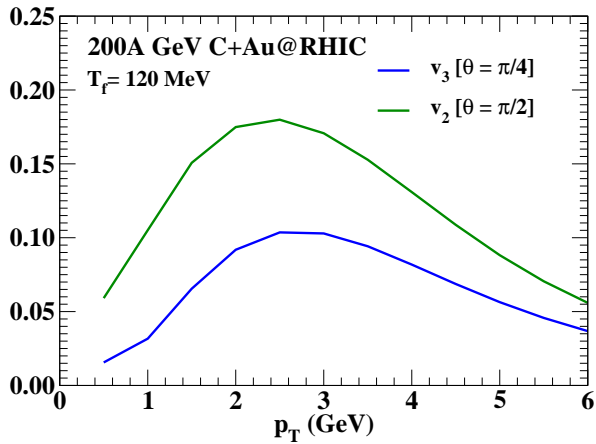}}
	\caption{{ (Color online) The thermal photon $v_3$ and $v_2$  from the clustered ${\rm C}$ and ${\rm Au}$  collisions at 200A GeV with orientation angles $\theta=\pi/4$ and $\pi/2$ respectively for $T_f=120$ MeV.}}
	\label{fig7}
\end{figure}

 It is to be noted that the value of $T_f$ is fixed for Au+Au collisions by reproducing the experimental data of charged particle spectra and anisotropic flow parameters.  We can not do the same for C+Au collisions due to the unavailability of experimental data. Therefore, we use the same $T_f$ for C+Au collisions as in Au+Au collisions.  A smaller $T_f$ would enhance the photon yield from the hadronic phase in the region $p_T < 1$ GeV and the larger $p_T$ part of the direct photon spectrum will remain unaltered. However, the photon anisotropic flow is expected to be larger for a smaller $T_f$. To show the effect of a relatively longer hadronic phase evolution on the photon flow observables we have calculated thermal photon $v_2$ and $v_3$ considering an arbitrary small $T_f$ $(=120$ MeV, see Fig.~\ref{fig7}) for orientation angles $\pi/2$ and $\pi/4$ respectively where, the flow coefficients are estimated to be largest. We see that $v_2$ and $v_3$ are enhanced for a smaller $T_f$, however, the qualitative difference between the photon anisotropic flow parameters  remains unchanged even for a significant change in the value of the freeze-out temperature. 
 A detailed study considering a larger formation time and dynamical freeze-out would be valuable in this regards and we postpone it for the future.\\

\begin{figure}[htbp!]
	{\includegraphics*[width=7.8cm,clip=true]{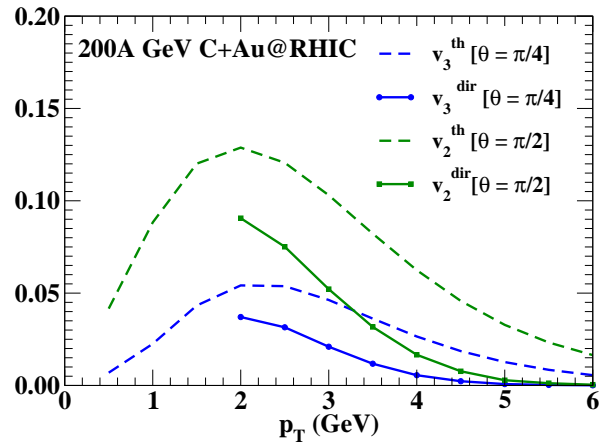}}
	\caption{(Color online) The direct photon $v_3$ and $v_2$  from the clustered ${\rm C}$ and ${\rm Au}$  collisions at 200A GeV with orientation angles $\theta=\pi/4$ and $\pi/2$ respectively.}
	\label{fig8}
\end{figure}

The anisotropic flow parameters of direct photons for the clustered configurations with the largest thermal photon $v_2$ and $v_3$ are shown in Fig.~\ref{fig8}. By `direct photons' we mean the combined contribution of thermal and the prompt photons (produced in initial hard scatterings). We find that the values of the anosotropic flow cofficients are substantially large even after inclusion of the prompt contribution (The prompt spectrum is generated in the region $p_T>2$ GeV (due to some constraints) using JETPHOX and as a result we could not show the anisotropic flow in the low $p_T$ region). The maximum value of direct photon $v_3$ and $v_2$ (at 2 GeV) are found to be about $70\%$ of their respective maximum thermal photon $v_3$ and $v_2$ values. It is to be noted that these values are comparable to the anisotropic flow parameter data obtained from the $0-20\%$ Au+Au and Pb+Pb collisions at RHIC and LHC energy respectively~\cite{Adare:2015lcd,Acharya:2018bdy}.  Thus, considering even a smooth initial condition, we can obtain a significantly large direct photon $v_2$ and $v_3$ depending upon the orientations of collision for the clustered carbon. We expect that the inclusion of initial state fluctuations would further increase the value of direct photon $v_2$ and $v_3$. On the other hand, photon $v_2$ and $v_3$ for unclustered carbon appear only due to initial state fluctuations and are expected to be significantly smaller than the clustered carbon.
Thus, the clustered carbon and gold collisions at the RHIC have the potential to produce considerably large anisotropic flow parameters of photons even for the most central events. Furthermore, the strong sensitivity of the anisotropic flow parameters to the initial geometry and early QGP evolution also can be valuable to understand the `direct photon puzzle'. Our main focus in the present calculation is only to estimate the effect of initial geometry on the production and anisotropic flow parameters of photons. A more realistic calculation, using an event-by-event (3+1) dimensional hydrodynamic framework, would be valuable to precisely explore the role of fluctuations and participant asymmetry on the evolution of thermodynamic quantities and anisotropic flow parameters of photons.

\subsection{Geometry (flow) correlation}

\begin{figure}[htbp!]
	{\includegraphics*[scale=0.50,clip=true]{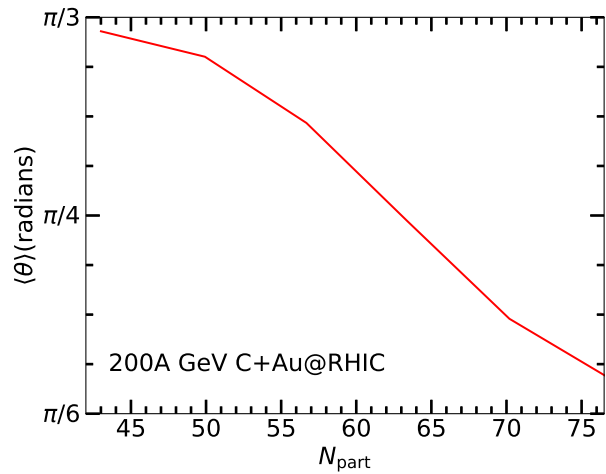}}
	\caption{(Color Online) Average orientation angle $\langle\theta\rangle$ as a function of $N_{\rm part}$ for $\alpha$-clustered ${\rm C}$ + ${\rm Au}$ collisions  at 200A GeV. }
	\label{fig9}
\end{figure}

To get an idea of the experimental signature, we simulate a large number of clustered C+Au events with different configurations and compare the results obtained from the collisions of unclustered carbon. Fig.~\ref{fig9} shows the average orientation angle $\langle\theta\rangle$ as a function of the number of participants in the collisions of clustered carbon, where and henceforth $\langle ... \rangle$ denotes an event average. We observe that with the increasing $N_{\rm part}$, the average angle reduces significantly. Thus, based on the multiplicity cut, it may be possible to select events with various configurations.

\begin{figure}[htbp!]
	{\includegraphics*[scale=0.50,clip=true]{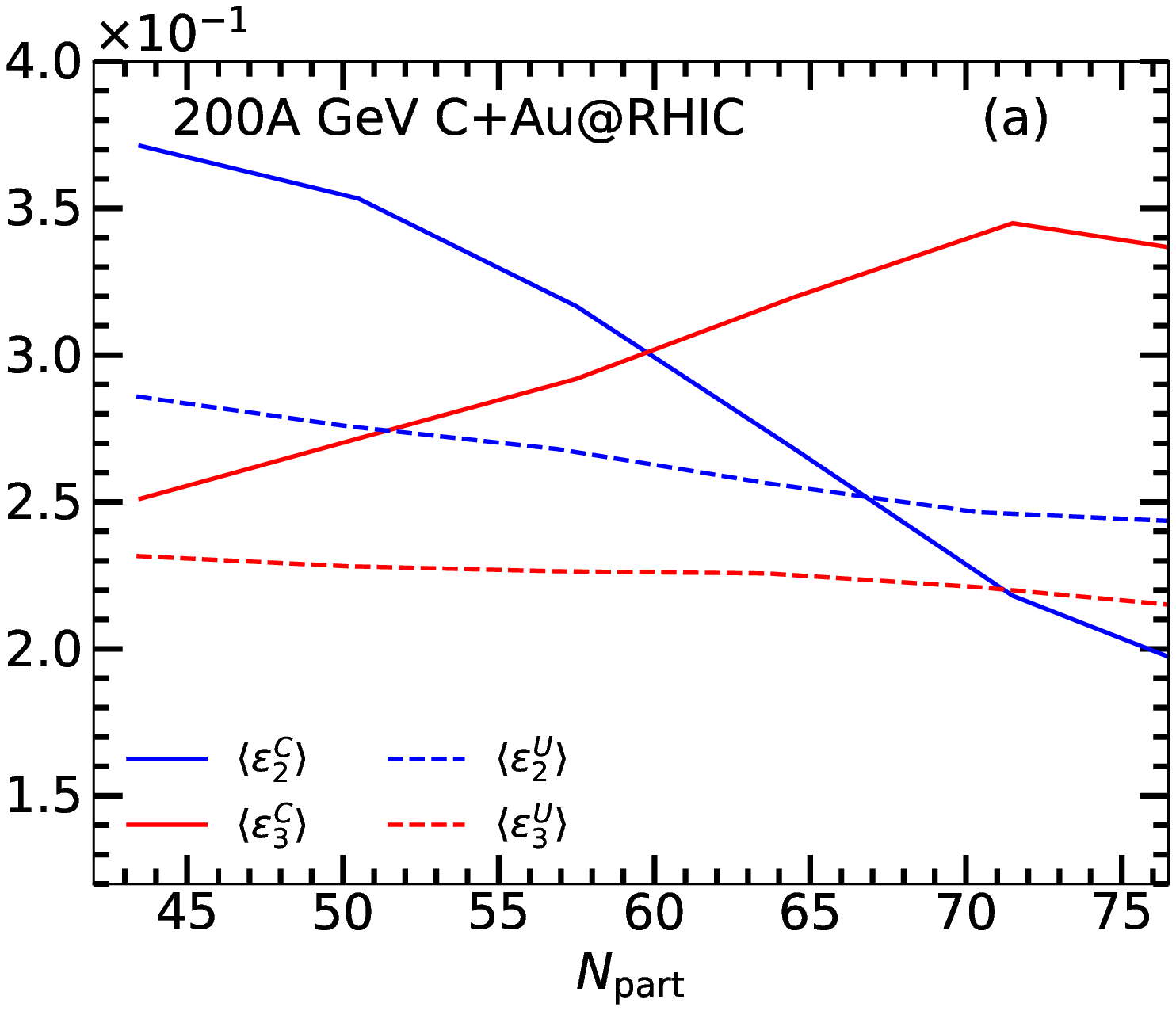}}
	{\includegraphics*[scale=0.52,clip=true]{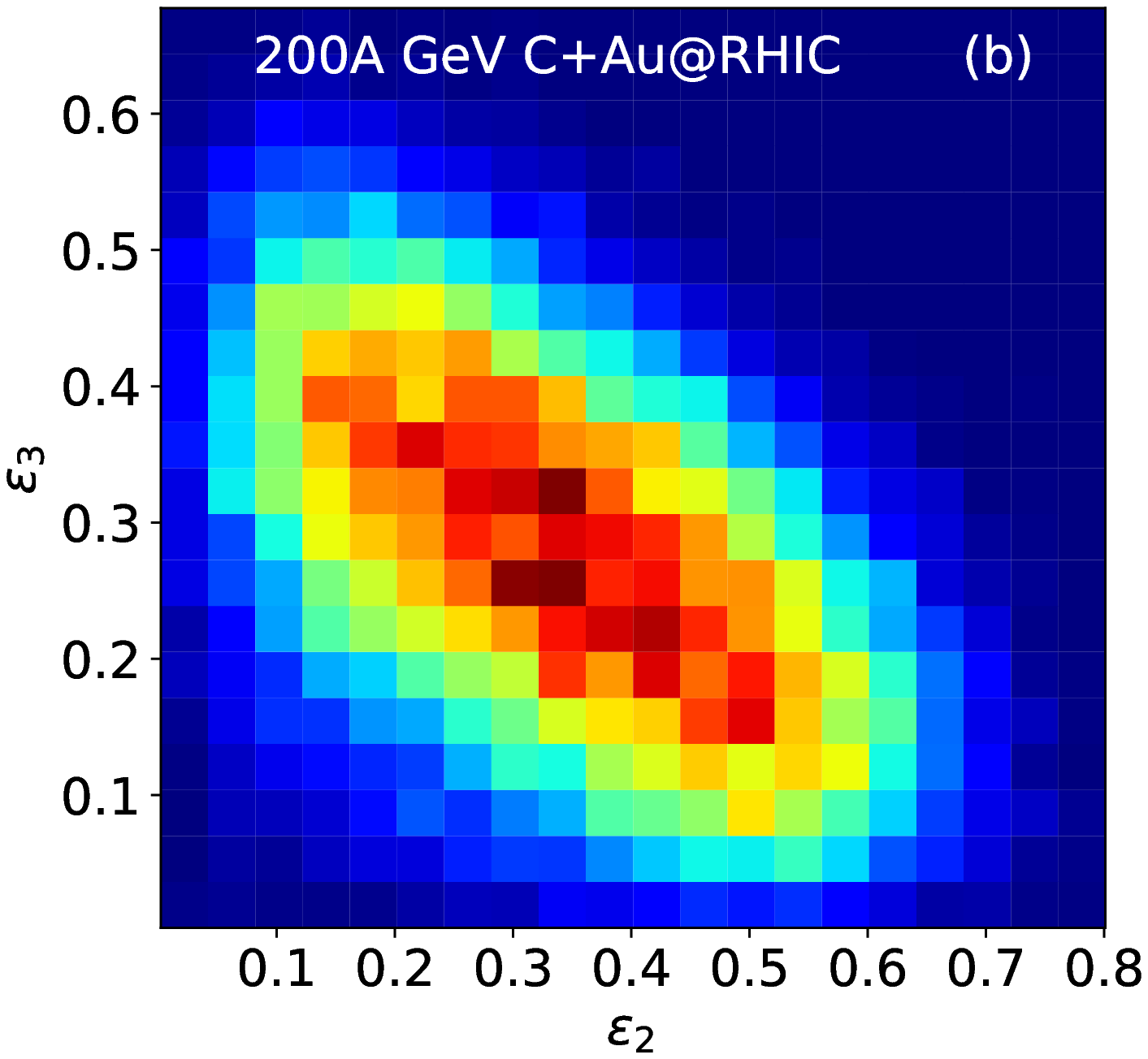}}
	{\includegraphics*[scale=0.50,clip=true]{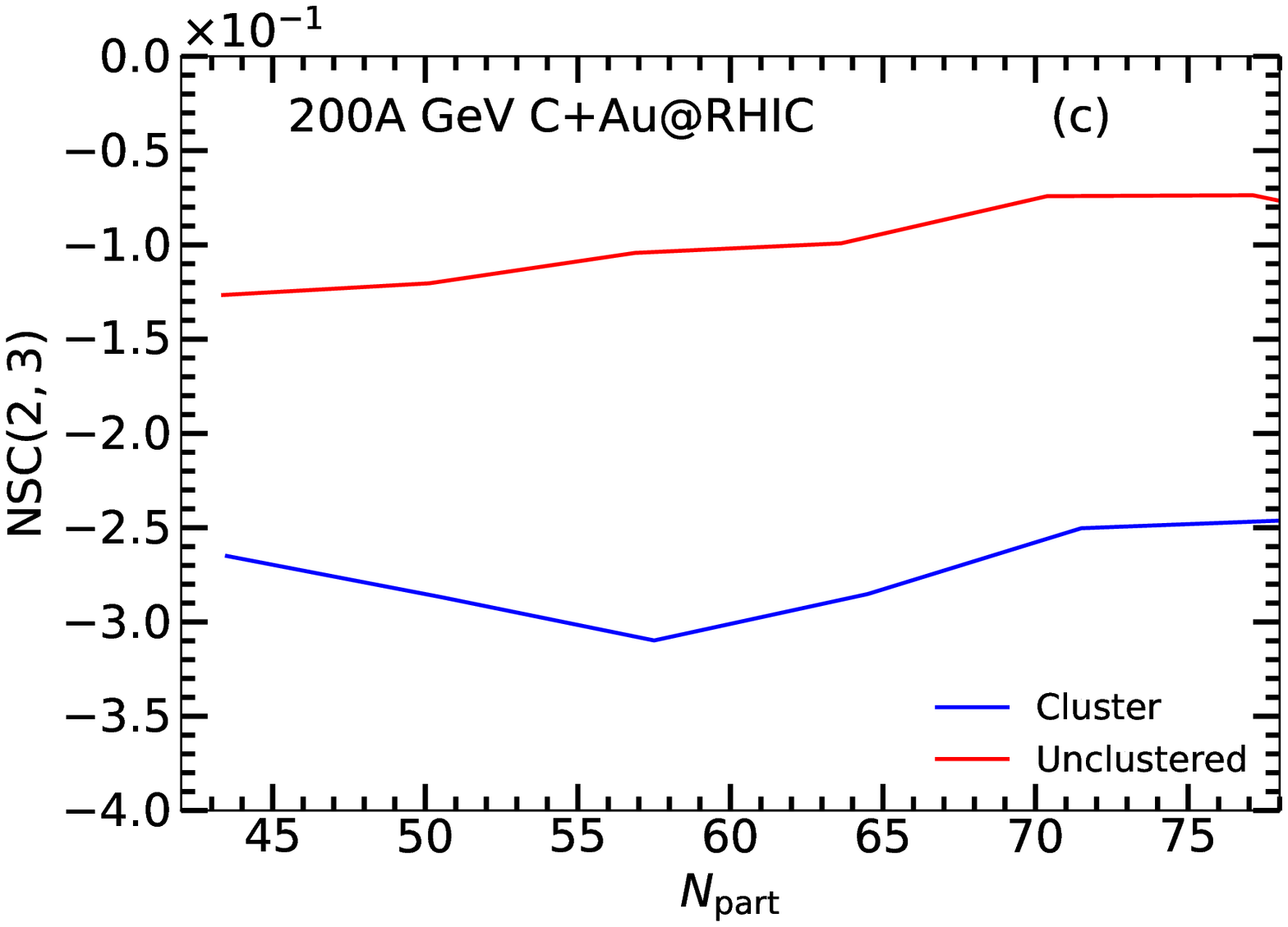}}
	\caption{(Color Online)  (a) The average  ellipticity and triangularity as a function of $N_{\rm part}$  for $\alpha$-clustered and unclustered ${\rm C}$+  ${\rm Au}$ collisions  at 200A GeV  (b) Event distribution in $\varepsilon_2$-$\varepsilon_3$ plane and  (c) The normalized symmetric cumulants NSC(2,3) as functions of $N_{\rm part}$ for unclustered and $\alpha$-clustered cases. The events are chosen with $N_{\rm part} > 40$.}
	\label{fig10}
\end{figure}

In Fig.~\ref{fig10}(a) we present  the $N_{\rm part}$ dependence of average eccentricity and triangularity in unclustered and $\alpha$-clustered  ${\rm C+Au}$ collisions at 200A GeV.  We consider the events having  $N_{\rm part} > 40$. The superscripts `U' and `C' stand for the unclustered and clustered carbon respectively. We see that $\langle \varepsilon_2^U \rangle$ and $\langle \varepsilon_3^U \rangle $ show similar behavior with increasing $N_{\rm part}$. Due to vanishing geometric deformation and smaller fluctuations at a larger $N_{\rm part}$, the average eccentricity decreases. We observe a different trend for clustered carbon. Due to the intrinsic triangular geometry of the clustered-carbon, $\langle \varepsilon_3^C \rangle$ increases with $N_{\rm part}$, whereas $\langle \varepsilon_2^C \rangle $ decreases. The difference becomes substantial at larger $N_{\rm part}$, which indicates an obvious anti-correlation between ellipticity and triangularity exists in $\alpha$-clustered ${\rm C}$ + ${\rm Au}$ collisions.

In Fig.~\ref{fig10}(b) we plot the event density distribution in $\varepsilon_2 - \varepsilon_3$ plane for the collisions of clustered carbon. To quantify the anti-correlation, we define a normalized symmetric cumulants coefficient NSC(n,m) as follows~\cite{Giacalone:2016afq},
\begin{eqnarray}
\rm NSC(n,m)=\frac{\langle{\varepsilon_n}^2{\varepsilon_m}^2\rangle-\langle{\varepsilon_n}^2\rangle \langle{\varepsilon_m}^2\rangle}{\langle\varepsilon_m^2\rangle\langle\varepsilon_n^2\rangle}.
\label{nsc}
\end{eqnarray} 
The NSC(2,3)  is expected to be similar to the quantity involving the anisotropic flow coefficients ($v_2$ and $v_3$) under a linear response assumption~\cite{Noronha-Hostler:2015dbi,Yan:2015jma}. In Fig.~\ref{fig10}(c), we show the normalized symmetric cumulants coefficient as function of $N_{\rm part}$ for both unclustered and $\alpha$-clustered cases. The negative values of  coefficients signify the anti-correlation relation between the two variables. We see a larger anti-correlation for the clustered case compared to the unclustered case. The coefficients for both the cases do not show significant $N_{\rm part}$ dependence. The NSC(2,3) for the clustered carbon is found to be three times larger than the NSC(2,3) obtained for the unclustered carbon. We find a small minimum of NSC(2,3) at $N_{\rm part} \sim 57$, where the values of the $\langle\varepsilon_2^C\rangle$ and $\langle\varepsilon_3^C\rangle$ are close to each other.  Exploring the large anti-correlation between ellipticity and triangularity (elliptic flow and triangular flow), we can detect the $\alpha$-cluster effects of carbon in ${\rm C + Au}$ collisions. It is to be noted that some earlier studies have explored the characteristics of the initial-state eccentricities and final hadron flow observables from the collisions of the clustered carbon and heavy-ions at various beam energies in the event-by-event framework~\cite{Broniowski:2013dia,Bozek:2014cva}. In principle, we expect that the anti-correlation behavior should hold for both photons and hadrons, since their flow observables basically arise from the initial geometry.

\section{SUMMARY}
We have calculated $p_T$ spectra and differential anisotropic flow coefficients of thermal photons for $\alpha$-clustered ${\rm C}$ and ${\rm Au}$ collisions at RHIC using a hydrodynamic model with the event-averaged smooth initial density distribution and compared the results with those obtained from unclustered carbon and gold collisions. Three different orientations of collision for the $\alpha$-clustered carbon have been considered which produce notably different initial spatial ellipticities and triangularities. 

 The slope of the thermal photon spectrum depends on the orientation of collision. A larger initial temperature for the orientation angle $\theta = \pi/2$ results in more high $p_T$ photons from the early stage of system evolution compared to the other configurations. However, we see no significant difference in the direct photon spectra for various configurations due to the large contribution of prompt photons in the region $p_T> 3$ GeV. We see significantly large thermal photon $v_2$ and $v_3$ from the configurations $\theta = \pi/2$ and $\theta = 0$, $\pi/4$ respectively, which are large and comparable to the experimental data of photon $v_2$ and $v_3$ from the mid-central 200A GeV Au+Au collisions at RHIC. On the other hand, photon $v_2$ and $v_3$ from the unclustered ${\rm C+Au}$ collisions are close to zero as the corresponding spatial anisotropies are vanishingly small for the system. We find that addition of the prompt photon contribution dilutes the anisotropic flow parameters but still their values are substantially larger in comparison to the theoretical estimate of anisotropic flow parameters of direct photons obtained for the mid-central Au+Au collisions. We have studied the individual contributions of the thermal photon $v_2$ and $v_3$ from QGP and hadronic phases and have found that these anisotropic flow coefficients are predominantly built during the QGP evolution. As the effect of viscosity which prolongs the evolution of the QGP is not included here, we expect that it could change the magnitude but not the qualitative nature of these photon observables. Thus, we find that anisotropic flow of photons from the clustered carbon are sensitive to the anisotropic nuclear distribution of ${\rm C}$ and can potentially  help us to understand the photon $v_n$ puzzle. A strong anti-correlation between ellipticity and triangularity is observed because of the exotic internal structure of $\alpha$-clustered carbon. With the aid of these special features of photon spectra and anisotropic flows, it could be a potential probe to detect the exotic internal structure of $\alpha$-clustered carbon in the future high energy ${\rm C}$ + ${\rm Au}$ experiment.

\begin{acknowledgments}
This work is supported by the National Natural Science Foundation of China under Grants No. 11835002, No. 11890710, No. 11890714, No. 11975079 and No. 11961131011, the Strategic Priority Research Program of Chinese Academy of Sciences under Grant No. XDB34030000, Shanghai Pujiang Program under Grant No. 19PJ1401400, and the Guangdong Major Project of Basic and Applied Basic Research under Grant No. 2020B0301030008.
\end{acknowledgments}

\vspace{1.50 cm}

\end{document}